\documentclass[11pt]{article}
\pdfoutput=1
\usepackage{amsmath,graphicx}
\usepackage{hyperref}
\usepackage[T1]{fontenc}
\usepackage{mathpazo}
\usepackage[font=small,labelfont=bf]{caption}
\usepackage{cancel}
\usepackage{cite}
\usepackage{multirow}

\newcommand{\beq}{\begin{eqnarray}}
\newcommand{\eeq}{\end{eqnarray}}
\def\simlt{\stackrel{<}{{}_\sim}}
\def\simgt{\stackrel{>}{{}_\sim}}

\usepackage[table,usenames,dvipsnames]{xcolor}
\usepackage{framed}
\usepackage{afterpage}

\title{\bf \color{red!80!black} Effective field theory and keV lines from dark matter}
\author{Rebecca Krall, Matthew Reece, and Thomas Roxlo\\
{\em Department of Physics, Harvard University, Cambridge, MA 02138, USA}}

\begin{document}
\maketitle

\begin{abstract}
We survey operators that can lead to a keV photon line from dark matter decay or annihilation. We are motivated in part by recent claims of an unexplained 3.5 keV line in galaxy clusters and in Andromeda, but our results could apply to any hypothetical line observed in this energy range. We find that given the amount of flux that is observable, explanations in terms of decay are more plausible than annihilation, at least if the annihilation is directly to Standard Model states rather than intermediate particles. The decay case can be explained by a scalar or pseudoscalar field coupling to photons suppressed by a scale not far below the reduced Planck mass, which can be taken as a tantalizing hint of high-scale physics. The scalar case is particularly interesting from the effective field theory viewpoint, and we discuss it at some length. Because of a quartically divergent mass correction, naturalness strongly suggests the theory should be cut off at or below the 1000 TeV scale. The most plausible such natural UV completion would involve supersymmetry. These bottom-up arguments reproduce expectations from top-down considerations of the physics of moduli. A keV line could also arise from the decay of a sterile neutrino, in which case a renormalizable UV completion exists and no direct inference about high-scale physics is possible.
\end{abstract}

\section{Introduction}
\label{sec:intro}

Searches for dark matter typically focus on WIMPs with a mass near the weak scale, but a wide range of masses merit consideration. For instance, the range of masses between a keV and several GeV is compatible with a number of interesting models, and astrophysical constraints exist from X-ray or gamma-ray observations~\cite{Essig:2013goa}. Recent claims of a possible 3.5 keV X-ray line observed in galaxy clusters and Andromeda~\cite{Bulbul:2014sua, Boyarsky:2014jta} have generated a great deal of interest in models with signals in the X-ray range~\cite{Ishida:2014dlp,Finkbeiner:2014sja,Higaki:2014zua,Jaeckel:2014qea,Lee2014}.

Our goal in this paper is to study the prospects for X-ray lines with energies of order a keV from an effective field theory perspective. If dark matter decays or annihilates directly to Standard Model states and produces such a line, there is a small set of operators that are interesting to consider. In fact, for any two-body final state below the two-electron threshold $2 m_e \approx 1~{\rm MeV}$, only photons and neutrinos can appear in the final state, significantly restricting the possible signatures. We will analyze the size of the operator needed to fit the 3.5 keV line signal. Even if this signal proves to not be a result of dark matter decay or annihilation, the resulting effective field theory analysis will carry over to any future line signal in this energy range. 

An earlier operator analysis of monochromatic photon lines in dark matter decay, not limited to the X-ray regime, appeared in Ref.~\cite{Gustafsson:2013gca}. It is complementary to ours, as it includes a wider set of decays that are available for higher-mass dark matter and discusses the line-to-continuum ratio in such cases. On the other hand, by specializing to the X-ray regime we have a definite observable flux to target and can make inferences about the scales that could suppress the operator. We also add a discussion of loops and naturalness, with implications for possible UV completions of the effective theory.

We first consider two-body decays of dark matter. If DM is bosonic, it decays to two photons. This requires that it is a scalar or pseudoscalar, because a massive spin-1 particle cannot decay to two massless photons due to the Landau-Yang theorem~\cite{Landau:1948kw, Yang:1950rg}. Thus, the interesting possibilities are a scalar $\phi$ or a pseudoscalar $a$ decaying through the operators:
\beq
\frac{\phi}{f} F_{\mu \nu}F^{\mu \nu}~~{\rm or}~~\frac{a}{f} F_{\mu \nu} {\tilde F}^{\mu \nu}.
\eeq
In the fermionic case, the final state must be a photon and a neutrino, and the most plausible option is a spin-$1/2$ fermion $\psi$ (which could be thought of as a sterile neutrino):
\beq
\frac{v}{f^2} \psi^\dagger \sigma_{\mu \nu} \nu F^{\mu \nu}
\eeq
Here $v$ is the Higgs VEV which has been put in because in the full SM, gauge invariance requires the neutrino field to appear through the operator $HL$. Alternatively, one could consider a similar operator with a spin-$3/2$ gravitino, though we will find that the required rate is difficult to achieve in this case. 

If dark matter {\em annihilates} to SM final states and produces a 3.5 keV line, the operator must be suppressed by a small scale, which is difficult to UV complete. Thus, the decaying interpretation seems more plausible, although more elaborate 2-body processes that do not involve {\em directly} annihilating to SM states are possible (see, e.g., ref.~\cite{Finkbeiner:2014sja}).

From the effective field theory perspective, the richest and least-studied case turns out to be scalar dark matter. We find that detectable line signals coincide with operators suppressed by scales of order the reduced Planck mass $M_{\rm Pl} \approx 2.4 \times 10^{18}~{\rm GeV}$. Like all scalar fields, such dark matter would have a severe naturalness problem unless divergences are regulated by new physics. We argue that in this case, the cutoff should be below about $\sqrt{m_\phi M_{\rm Pl}} \sim {\rm PeV}$, and that the most plausible new physics leading to a consistent and natural theory is supersymmetry. It is very interesting that a line signal in the X-ray energy range could be correlated with supersymmetry at a scale that can be probed by, if not current colliders, at least possible future ones.

\section{Decaying Scalar Dark Matter}
\label{sec:scalar}

\subsection{Decays in the effective theory}

We consider an effective interaction for a scalar field $\phi$ of mass $\approx 7$ keV coupled to photons:
\beq
{\cal L}_{\rm int} = \frac{\phi}{f} F_{\mu \nu}F^{\mu \nu}. 
\label{eq:scalarLint}
\eeq
This leads to a decay width
\beq
\Gamma(\phi \to \gamma \gamma) = \frac{m_\phi^3}{4 \pi f^2}.
\eeq
The range of sterile neutrino mixing angle estimates in Ref.~\cite{Boyarsky:2014jta} correspond to a lifetime for a fermion decaying to $\gamma\nu$ of between $2.0 \times 10^{27}$ and $1.9 \times 10^{28}$ seconds. The value of the mixing angle quoted in the abstract of Ref.~\cite{Bulbul:2014sua} corresponds to a lifetime of $5.7 \times 10^{27}$ seconds. For a decay to two photons, the estimated lifetime should be twice as long. We will thus take our goal to be a rate of
\beq
\Gamma(\phi \to \gamma \gamma) \approx \frac{1}{1.2 \times 10^{28}~{\rm s}} \approx 5 \times 10^{-53}~{\rm GeV},
\eeq
keeping in mind that the rate is uncertain to within about a factor of 2 around this value. Given a measured mass $m_\phi \approx 7.1~{\rm keV}$, we find that this rate requires a decay constant $f$ of order the reduced Planck mass:
\beq
f \approx 7 \times 10^{17}~{\rm GeV},
\eeq
to within about 50\% accuracy.

This is a very interesting result, since scalar fields with couplings suppressed by scales of order $M_{\rm Pl}$ are well-motivated: they arise in the form of {\em moduli} in string theory constructions, which tend to be generated in large numbers cosmologically (to the extent of posing a problem)~\cite{Coughlan:1983ci,Ellis:1986zt,deCarlos:1993jw,Banks:1993en}. In fact, the possibility of X-rays arising from decaying moduli has been noted before and constitutes an important constraint on light moduli fields~\cite{Kawasaki:1997ah,Hashiba:1997rp,Kusenko:2012ch}. Before turning to such a top-down interpretation, we will first consider what we can learn about such a scalar field from a purely bottom-up effective field theory perspective.

\subsection{Loops and Naturalness in the Effective Theory}


\begin{figure}[h]
\begin{center}
\includegraphics[width=0.5\textwidth]{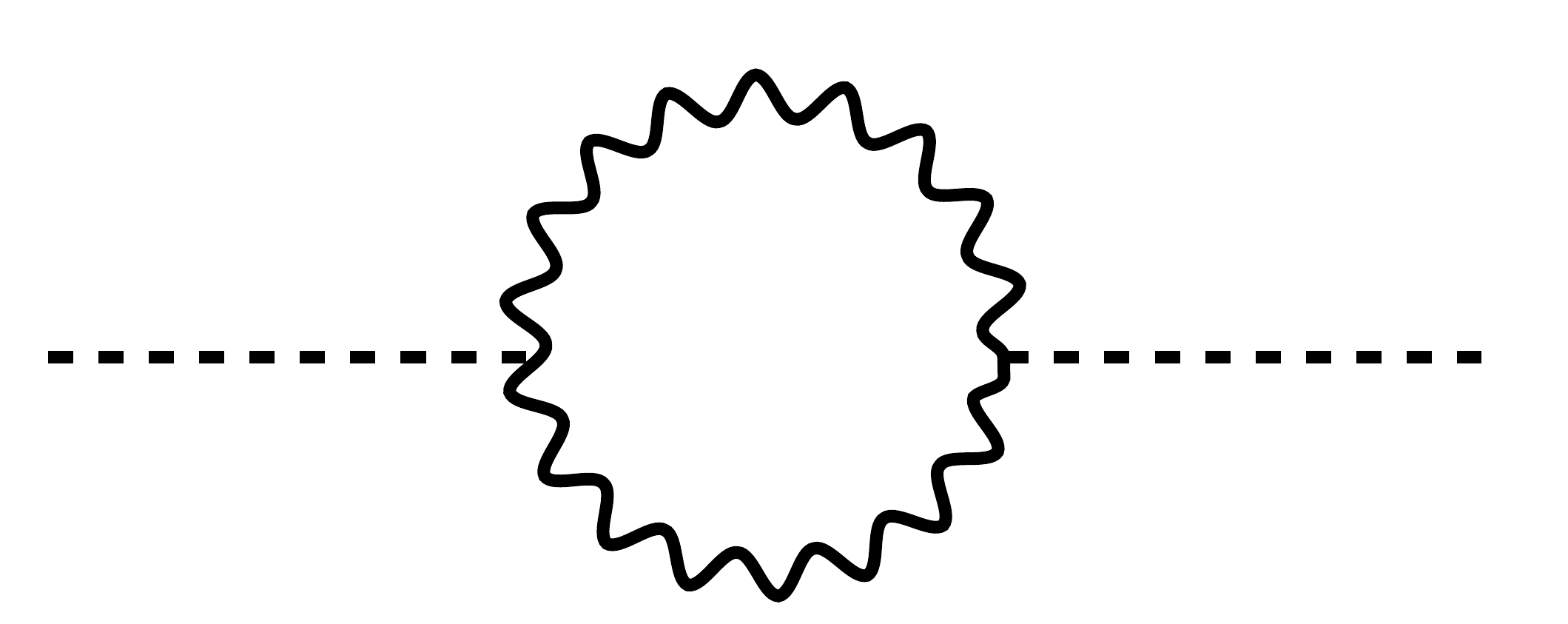}
\end{center}
\caption{Quartically-divergent loop correction to the scalar mass given the interaction in eq.~\ref{eq:scalarLint}.
} \label{fig:scalardivergentloop}
\end{figure}

There is an interesting naturalness constraint in this theory. The mass of $\phi$, like all scalar masses, is very sensitive to renormalization effects, such as the loop correction in fig.~\ref{fig:scalardivergentloop}. Because the coupling is $1/f$, we should expect just on dimensional analysis grounds that if loops are cut off at a scale $\Lambda$ we generate quartically-divergent corrections. Calculating the diagram with a simple cutoff gives
\beq
\delta m_\phi^2 \approx \frac{3}{8 \pi^2} \frac{\Lambda^4}{f^2}.
\eeq
The precise coefficient is cutoff-dependent, which in more physical language means it is dependent on the unknown short-distance physics that softens the divergent loop. Nonetheless, this estimate should capture the effect at an order-of-magnitude level. 

Naturalness is the expectation that this correction to the mass of $\phi$ should not be much larger than the actual observed mass of $\phi$, which would require a delicate cancelation. In this context, it tells us that
\beq
\Lambda \simlt \left(\frac{8 \pi^2}{3}\right)^{1/4} \sqrt{m_\phi f} \approx 4 \times 10^6~{\rm GeV}.
\eeq
Any natural UV completion of this decaying scalar theory has a cutoff at or below a few thousand TeV (that is, a few PeV) that softens the divergent photon loop. We will consider what such a UV completion may be below. In any case, it is appealing that the low energy scale corresponding to the X-ray line and the very high energy scale implied by its rate have conspired to imply new physics at the PeV scale. This is almost within reach of, if not current colliders, at least technologically plausible future colliders.

As in all naturalness arguments, we could always imagine that there is a single fine-tuning in the theory that produces a light physical $\phi$ mass but that the cutoff of the theory remains at the scale $f$. In this case, we would expect that the scale $f$ suppresses generic higher-dimension operators in the theory. For example, we should also add to the Lagrangian
\beq
{\cal L}_{\rm int}' = \frac{c}{f^2} \phi^2 F_{\mu \nu}F^{\mu \nu},
\label{eq:newscalarterm}
\eeq 
with $c$ an order-one number. Generic such operators suppressed by $f$ will contribute comparable amounts to the scattering amplitude (e.g. for $\phi \gamma \to \phi\gamma$) at center-of-mass energies $\sim f$, so the effective theory breaks down at that scale and it is the correct cutoff.

\begin{figure}[h]
\begin{center}
\includegraphics[width=0.95\textwidth]{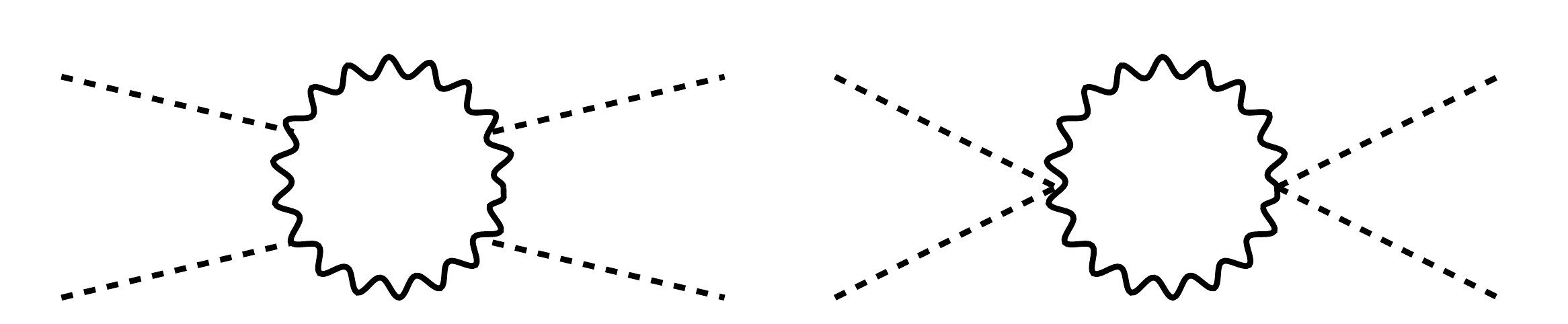}
\end{center}
\caption{Quartically-divergent loop corrections $\delta \lambda \sim \frac{1}{16\pi^2} \frac{\Lambda^4}{f^4}$ to the scalar self-interaction given the interactions in eq.~\ref{eq:scalarLint} and~\ref{eq:newscalarterm}.
} \label{fig:scalardivergent4pt}
\end{figure}

Even after we tune the $\phi$ mass to be light, loops generating the $\phi \phi \to \phi \phi$ scattering amplitude pose a further phenomenological problem. Diagrams contributing to this amplitude are shown in fig.~\ref{fig:scalardivergent4pt} and are also quartically divergent. Although we did not initially write down a $\phi^4$ term in the potential, it is allowed by all symmetries, and if the cutoff of the theory is $\Lambda \sim f$ these loop diagrams will generate $\frac{\lambda}{4!} \phi^4$ with $\lambda \simgt \frac{1}{8\pi^2}$. Such a value of $\lambda$ may seem innocuous, but self-interactions of dark matter have observable effects, like isotropizing the velocity distribution of dark matter in the dense cores of galaxies and galaxy clusters and thus producing more spherical halos. Such halo shape constraints, as well as dynamics of the Bullet Cluster merger, have been used to set bounds on dark matter self-interactions~\cite{MiraldaEscude:2000qt,Buote:2002wd,Markevitch:2003at,Randall:2007ph,Rocha:2012jg,Peter:2012jh}. The current best estimates are that the allowed cross sections are
\beq
\frac{\sigma}{m_{\rm DM}} \simlt 0.1~{\rm barn}/{\rm GeV},
\eeq
which in our case means
\beq
\sigma \simlt 7 \times 10^{-31}~{\rm cm}^2 \approx 2 \times 10^{-3}~{\rm GeV}^{-2}.
\eeq
This would compare to our estimate
\beq
\sigma_{\rm est} \sim \frac{1}{32 \pi s} \left|\lambda\right|^2 \sim 8 \times 10^3~{\rm GeV}^{-2},
\eeq
using $\lambda = \frac{1}{8\pi^2}$ and the fact that $s \approx 4 m_\phi^2$ for nonrelativistic scattering. As a result, tuning the mass to be small is not enough. Data requires that the self-interaction cross-section be significantly suppressed relative to the one-loop effective field theory estimate if the theory is cut off at $\Lambda \approx f$. A cutoff two orders of magnitude below $f$ would safely avoid this problem while still requiring a large fine-tuning in the mass. Nonetheless, this secondary argument strengthens the naturalness point of view that the theory breaks down below the scale $f$, and we find the case for a breakdown of the theory at or below a PeV to be compelling.

\subsection{A composite UV completion}

The naturalness argument implies that some new physics should come in at the PeV scale to cut off the quartic divergence and make the $\phi$ mass natural. We will now explore some possibilities for what this PeV-scale physics can be. As in the more familiar hierarchy problem arising from the Higgs boson mass, the physics that explains why $\phi$ is light could arise from compositeness or from supersymmetry. Let us first address the less appealing case of compositeness. We could begin with a Dirac fermion $\chi$ charged under some SU($N$) gauge group, coupling to photons through the dimension-7 Rayleigh  operator $\frac{1}{\Lambda_R^3} \chi{\bar \chi} F_{\mu \nu} F^{\mu \nu}$. Such an operator can be readily UV completed by loops of electrically charged particles that could also carry SU($N$) quantum numbers~\cite{Weiner:2012gm}. We could then imagine that SU($N$) confines, binding the $\chi$ particles into a composite scalar $\phi$ with a mass set by the compositeness scale, i.e. $\Lambda_{\rm comp} \sim m_\phi \sim 10~{\rm keV}$. In that case, we expect that
\beq
\frac{1}{f} \sim \frac{\Lambda_{\rm comp}^2}{\Lambda_R^3} \Rightarrow \Lambda_R \sim \left(m_\phi^2 f\right)^{1/3} \approx 300~{\rm GeV}.
\eeq
There are several reasons that this seems unlikely: first, it would require new charged particles at the weak scale, also charged under a hidden gauge group, that we have not yet seen. Second, it requires a light stable scalar in the composite sector, built out of fermionic constituents, whereas confinement typically favors pseudoscalars as the lightest states. (Glueballs are an exception, but more difficult to UV complete.) Third, it involves a gauge group that confines at the keV scale, which could pose cosmological difficulties (though our favored alternative is also cosmologically problematic). Finally, if nature has handed us a scale like $7 \times 10^{17}~{\rm GeV}$ suppressing an operator, it would be disappointing if it is completed by such mundane physics. Instead, the obvious temptation is to view this very large scale as a fundamental one, near the Planck scale for a good reason rather than an accidental one.

\subsection{Supersymmetry as a cutoff}

We now wish to explore the possibility that the field $\phi$ is not composite, and the large scale $f$ is related to fundamental physics (like the Planck scale or string scale). Large fundamental scales suppressing couplings are typical of a class of scalar fields called moduli. For the remainder of this section, we will explore the possibility that decaying scalar moduli fields give rise to an X-ray line. In this case, the quartic divergence at the PeV scale can be regulated by supersymmetry.

Scalar moduli fields arise, for instance, from modes of the graviton polarized along internal dimensions in string compactifications. Being part of the gravitational sector of the underlying theory, they couple with a strength related to that of gravity, so it would not be a surprise to find them with $f \approx 7 \times 10^{17}~{\rm GeV} \approx 0.3 M_{\rm Pl}$. Moduli fields typically have a mass that is at most of order the gravitino mass $m_{3/2}$, which is a measure of how badly supersymmetry is broken~\cite{deCarlos:1993jw}. Although they may obtain a mass in a supersymmetric manner, which is sometimes moderately large compared to $m_{3/2}$~\cite{Kachru:2003aw}, raising their mass well above the gravitino mass generally requires fine-tuning~\cite{Fan:2011ua}. In the simplest case (though see below for an exception in LARGE volume models), moduli masses are of order $m_{3/2}$. As a result, we expect that if a modulus field is found with a mass of 7 keV, the gravitino mass $m_{3/2} = \frac{F}{\sqrt{3} M_{\rm Pl}}$ is somewhere between about 100 eV and 10 keV, corresponding to a SUSY breaking scale
\beq
\sqrt{F} = \left(\sqrt{3} M_{\rm Pl} m_{3/2}\right)^{1/2} \approx 600~{\rm TeV}~{\rm to}~6~{\rm PeV}. \label{eq:SUSYbreaking}
\eeq
It is no accident that this is close to the cutoff we estimated above, since $f \sim M_{\rm Pl}$ and we have assumed $m_{3/2} \sim m_\phi$. But in making this estimate, we are using input from our theoretical expectations about moduli. It is interesting to try to reproduce such an estimate from a bottom-up, effective field theory point of view. 

In a supersymmetric theory, the photon loop in fig.~\ref{fig:scalardivergentloop} is supplemented by a photino loop. Supersymmetry is not powerful enough to completely remove ultraviolet divergences in a nonrenormalizable effective theory. It typically reduces the degree of a power divergence by 2, as in the familiar case when adding a stop loop to the top loop transforms a quadratically divergent Higgs mass into a logarithmically divergent one. A similar phenomenon occurs in this case: the quartically divergent loop correction to the $\phi$ mass becomes quadratically divergent in the presence of supersymmetry. We can read off the dependence from a general result on the behavior of the one-loop K\"ahler potential in nonrenormalizable theories~\cite{Brignole:2000kg}, which contains a term
\beq
\Delta K_{\rm 1-loop} \supset - \frac{\Lambda^2}{16 \pi^2} \log \left(\frac{h+ h^\dagger}{2}\right)
\label{eq:Kahler1loop}
\eeq
where $h$ is the holomorphic gauge kinetic function, i.e. ${\cal L} \supset \left(\int d^2\theta \frac{1}{4} h {\cal W}^\alpha {\cal W}_\alpha + {\rm h.c.}\right)$. If SUSY breaking generates a photino mass $m_\lambda$ at a scale $M_{\rm messenger}$, then below this scale we can view the mass as an effective SUSY-breaking spurion in the gauge kinetic function, $h \supset \frac{1}{g^2} m_\lambda \theta^2$. In this spurion treatment we must take $\Lambda \approx M_{\rm messenger}$, because higher momentum running through the loop will resolve the underlying SUSY-breaking dynamics and invalidate the spurion approach. Thus, in this approach we revise our estimate of the correction to the $\phi$ mass to be:
\beq
\delta m_\phi^2 \sim \frac{m_\lambda^2 M_{\rm messenger}^2}{16 \pi^2 f^2}.
\eeq
In the case of gauge-mediated supersymmetry breaking, we would have $m_\lambda \sim \frac{\alpha}{\pi} F/M_{\rm messenger}$ and so by imposing $\delta m_\phi^2 \simlt m_\phi^2$ we get a naturalness bound on the SUSY-breaking scale $\sqrt{F}$:
\beq
\sqrt{F} \simlt \sqrt{\frac{4\pi^2}{\alpha} f m_\phi} \approx 100~{\rm PeV} \Rightarrow m_{3/2} \simlt 5~{\rm MeV}.
\eeq
This is related to the estimate in eq.~\ref{eq:SUSYbreaking}, differing numerically because of weak coupling factors. Essentially, it says that it is natural for the moduli to be a few orders of magnitude lighter than the gravitino, assuming only that they interact with photons. In complete models we expect eq.~\ref{eq:SUSYbreaking} to be a better guide to the relevant scale, but it is interesting that we can get close to it from a bottom-up perspective.

Another possibility is that moduli can have mass well {\em below} the gravitino mass. Although this seems surprising from the point of view of ``generic'' effective supergravity theories, it is possible without fine-tuning. It occurs in the context of the LARGE volume string compactification models~\cite{Balasubramanian:2005zx,Conlon:2005ki}, where no-scale structure appears in the effective theory. Such no-scale structure could in fact be very common in the string landscape. In these scenarios, there is an exponentially large dimensionless number ${\cal V}$, the volume of the internal geometry in string units, which plays a key role. The string scale is $M_{\rm Pl}/\sqrt{\cal V}$, the gravitino mass is $m_{3/2} \sim M_{\rm Pl}/{\cal V}$ (with most of the moduli masses nearby), and the K\"ahler modulus controlling the overall volume is lighter still, with $m_\phi \sim M_{\rm Pl}/{\cal V}^{3/2}$. Our analysis of the loop corrections to the $\phi$ mass above approximately carries through in this case, except that we should cut off the divergent loop at $\Lambda \approx M_{\rm string}$ rather than at a messenger scale. The result, using the fact that the gaugino mass should be no larger than the gravitino mass, is a correction
\beq
\delta m_\phi^2 \sim \frac{m_\lambda^2 M_{\rm string}^2}{16 \pi^2 f^2} \simlt \frac{m_{3/2}^2 M_{\rm string}^2}{16 \pi^2 M_{\rm Pl}^2} \sim \left(\frac{M_{\rm Pl}}{4 \pi {\cal V}^{3/2}}\right)^2,
\eeq
smaller than or at most of order the mass scale of the light volume modulus. In other words, there is a consistent power-counting in the small parameter ${\cal V}^{-1/2}$. The possibility of X-ray signals from decays of the overall volume modulus in these scenarios was discussed in Ref.~\cite{Conlon:2007gk}, albeit in the context of energies closer to 100 keV to 1 MeV. If the mass of the volume modulus is at 7 keV, the corresponding gravitino mass is in the vicinity of 500 GeV. In this case, the new physics needed to cut off the quartically divergent loop correction would enter at scales well below the necessary energy needed to restore naturalness.

\subsection{Further implications of moduli}

Large decay constants imply that moduli fields can take large values, which leads to cosmological difficulties. The large range in field space is also associated with additional interactions of $\phi$. In concrete constructions, one finds that they often have noncanonical K\"ahler potentials. In fact, this is guaranteed by basic consistency requirements of field theory. If our gauge kinetic term is
\beq
{\cal L}_{\rm gauge} = \left(-\frac{1}{4} + \frac{\phi}{f}\right) F_{\mu \nu} F^{\mu \nu},
\eeq
then something must prevent us from considering field values $\phi > f/4$, which would alter the sign of the kinetic term, producing ghosts and catastrophically destabilizing the vacuum. One way that this can happen is if such distances are infinitely far away in field space. We could achieve this with a very mild divergence if the actual kinetic term for $\phi$ is
\beq
{\cal L}_{\rm kin} = \frac{f^2}{2\left(f - 4 \phi\right)^2} \partial_\mu \phi \partial^\mu \phi.
\eeq
Thus, the choice $\phi = f (1/4 - \epsilon)$ lies a distance $\sim f \log\left|\epsilon\right|$ away in field space, so it is physically impossible to reach $\epsilon = 0$ and we never encounter ghosts in the theory. This is precisely what happens in string theory, where the limit $\phi \to f/4$ also typically coincides with the emergence of a tower of light modes that should be included in the effective theory~\cite{Ooguri:2006in}. Providing an infinite distance in field space to avoid ghosts is also the role of the logarithm in the 1-loop K\"ahler correction in eq.~\ref{eq:Kahler1loop}.

In the early universe, the modulus field $\phi$ will typically take on values of order $f$, oscillating in its potential once the Hubble expansion rate drops to $m_\phi$. A modulus field with a mass of 7 keV is stable on long timescales and poses an overabundance problem. The best bet for solving such a problem is probably a late period of inflation~\cite{Randall:1994fr,LythStewart,deGouvea:1997tn}, which produces enough entropy to dilute the moduli. Intriguingly, in a particular realization of this idea studied by one of us with a saxion driving thermal inflation the scale $\sqrt{F} \sim {\rm PeV}$, with moduli near a keV, is preferred~\cite{Fan:2011ua}.

As we have mentioned, the modulus mass is usually near the gravitino mass. Moduli would be inaccessible at colliders (just as gravitons are), but the gravitino could arise as a decay product. The lifetime of a neutralino decaying to a gravitino is
\beq
c\tau \approx 6~{\rm cm} \left(\frac{\sqrt{F}}{500~{\rm TeV}}\right)^4 \left(\frac{100~{\rm GeV}}{m_{{\tilde\chi}^0}}\right)^5,
\eeq
ranging from 10 cm to 1000 km as the supersymmetry breaking scale $\sqrt{F}$ varies in the range in eq.~\ref{eq:SUSYbreaking}. Thus, the lower end of this range---motivated if the modulus is stabilized supersymmetrically and is moderately heavy compared to the gravitino---could offer spectacular collider signatures. Macroscopically displaced vertices  could open up a new arena of precision measurement~\cite{Kawagoe:2003jv,Chang:2009sv,Meade:2010ji,Park:2011vw}.

\section{Other Dark Matter Decay Scenarios}

\subsection{Decaying pseudoscalar}

The case of a pseudoscalar $a$, or axion-like particle, has obvious similarities to the scalar case we have just discussed:
\beq
{\cal L}_{\rm int} = \frac{a}{f} F_{\mu \nu} {\tilde F}^{\mu \nu}.
\eeq
In this case, we find $\Gamma = m_\phi^3/(4 \pi f^2)$ and thus\footnote{It is no accident that this pseudoscalar result matches the scalar case: a supersymmetric coupling $\int d^2\theta \frac{S}{f} {\cal W}_\alpha {\cal W}^\alpha$ of a chiral superfield $S$ to a gauge field strength couples the real scalar part of $S$ to $F^2$ and the imaginary scalar part of $S$ to $F {\tilde F}$. SUSY requires that the real and imaginary scalars have equal decay widths, relating the two computations.}
\beq
f \approx 7 \times 10^{17}~{\rm GeV}.
\eeq
Note that, although it is possible for a scalar modulus to appear with a coupling that is simply Planck-suppressed, it is generally the case that an axion field will appear with a coefficient of order $\frac{\alpha}{4\pi F_a}$ where $F_a$ is the axion field range~\cite{Svrcek:2006yi}. Thus, we should associate the estimate above with a scale of order $F_a \sim 4 \times 10^{14}~{\rm GeV}$. Such a scale could be related to important high scale physics like the string scale. Because the axion-like particle interpretation of the keV line has been discussed already in refs.~\cite{Higaki:2014zua,Jaeckel:2014qea,Lee2014}, we will not dwell on it in detail, save to make a few comments on its relation to the scalar case.

Axion fields have a shift symmetry, which is broken by their mass (originating, for instance, from instantons) but not by the coupling to photons. Thus, unlike the scalar case, we will not find any one-loop divergences that lead us to naturalness arguments in effective field theory for a higher scale. However, like the scalar case, axions are cosmologically problematic because the fields tend to dominate the energy density of the universe. This suggests that some physics dilutes the axion field after it has started oscillating in its potential; such dilution must happen at late times and at temperatures below about $\sqrt{m_a M_{\rm Pl}}$, hinting at a scale for new physics. Furthermore, if the axion mass originates from instantons, we could expect new physics (like a confining gauge theory) at the scale of the geometric mean $\left(m_a F_a\right)^{1/2} \sim 50~{\rm TeV}$. Thus, some aspects of the expectations for new particle physics and new cosmology are the same for the pseudoscalar interpretation as for the scalar case, although the bottom-up arguments leading to these conclusions are weaker.

\subsection{Decaying fermion}

Suppose that our dark matter field is a fermion, $\psi$, decaying to a photon and neutrino:
\beq
{\cal L}_{\rm int} = \frac{v}{f^2} \psi^\dagger \sigma_{\mu \nu} \nu F^{\mu \nu} + {\rm h.c.}
\label{eq:fermionint}
\eeq
We have included a factor of $v$ in the normalization because in the full Standard Model SU(2)$_L$ invariance requires that the Higgs field appear. Given this interaction, we find a decay width
\beq
\Gamma(\psi \to \gamma \nu) = \frac{v^2 m_\psi^3}{2 \pi f^4}.
\eeq
To achieve a lifetime of $\sim 6 \times 10^{27}~{\rm s}$, we need a scale
\beq
f \approx 1.3 \times 10^{10}~{\rm GeV}.
\label{eq:fermionscale}
\eeq
The interaction in eq.~\ref{eq:fermionint} is not the only term we can write involving the fermion $\psi$ and the neutrino; we can also consider a mass mixing between the two. In fact, as is well-known in the sterile neutrino context, such a mass mixing will lead to a dipole transition of the sort we are considering~\cite{Pal:1981rm,Kusenko:2009up}.

In the sterile neutrino context, we can generate eq.~\ref{eq:fermionint} beginning from a renormalizable Lagrangian,
\beq
{\cal L} \supset \frac{1}{2} m_\psi \psi \psi + y H L \psi,
\eeq
with $H = \frac{1}{\sqrt{2}} v + \ldots$ the Higgs boson. This leads to a mixing angle between the sterile neutrino $\psi$ and the Standard Model neutrino: $\sin \theta = \frac{y v}{\sqrt{2} m_\psi} \approx 2.5 \times 10^7 y$. Given the small mixing angle needed to fit the X-ray line data as quoted in refs.~\cite{Bulbul:2014sua,Boyarsky:2014jta}, this means that the Yukawa coupling $y$ is truly tiny: $\approx 1.6 \times 10^{-13}$. In this case, the largeness of our estimate for $f$ in eq.~\ref{eq:fermionscale} has nothing to do with actual high energies; it is a sign of a very small coupling. The dipole operator is induced through a $W$ loop which leads to a decay rate $\Gamma = \frac{9}{1024 \pi^4} \alpha y^2 m_\psi^3/v^2$~\cite{Pal:1981rm,Kusenko:2009up}. This produces $f \approx 3.9 \pi v/\sqrt{e y}$, consistent with our estimate above.

The Lagrangian above generates a contribution to the light neutrino mass of $\frac{y^2 v^2}{2 m_\psi} \approx 10^{-7}~{\rm eV}$, much smaller than the actual neutrino masses. This is perfectly consistent; it simply means that the keV sterile neutrino we have here plays no significant role in explaining neutrino masses. (A wide range of such sterile neutrino models and phenomenology have been studied; see recent reviews~\cite{Kusenko:2009up,Boyarsky:2009ix,Abazajian:2012ys}.)  Ordinarily we consider right-handed neutrinos and the seesaw mechanism to avoid postulating extremely tiny Yukawas, but here we seem to have one anyway. Various flavor symmetries could explain its smallness; see, for instance, the recent model in ref.~\cite{Ishida:2013mva,Ishida:2014dlp}. From our viewpoint, explaining the X-ray line in terms of a sterile neutrino is less compelling than explaining it with a modulus or an axion, which offer clearer connections to physics at high scales. In the sterile neutrino case, we have a simple renormalizable UV completion, so there is no unambiguous guidepost toward further new physics. 

\subsection{Decaying gravitino}

In general, a gravitino couples to the supercurrent, so we expect that it decays to particles and their superpartners, e.g. a photon and a neutralino. However, in the presence of lepton number violating interactions like $W_{BRPV} = \sum_i \mu_i L_i H_u$, neutrinos and neutralinos can mix. Beginning with the coupling of the gravitino $\psi_\mu$ to the photino $\lambda$ and the photon,
\beq
{\cal L} \supset -\frac{i}{8 M_{\rm Pl}} {\overline \psi}_\mu \left[\gamma^\nu, \gamma^\rho\right]\gamma^\mu \lambda F_{\nu \rho},
\eeq
the resulting decay rate will be proportional to the mixing matrix element $U_{{\tilde \gamma} \nu}$ between the photino and the neutrino~\cite{Takayama:2000uz}:
\beq
\Gamma({\tilde G} \to \gamma \nu) \approx \frac{1}{32 \pi} \left|U_{{\tilde \gamma} \nu}\right|^2 \frac{m_{3/2}^3}{M_{\rm Pl}^2}.
\eeq
Apart from the mixing factor this is parametrically very similar to the decay rate of moduli. But this is problematic for a gravitino interpretation: to obtain the right rate, we need $U_{{\tilde \gamma} \nu}$ to be an order-one number. However, we expect that in any realistic model of $R$-parity violating interactions, this mixing will be suppressed by a small parameter characterizing lepton number violation. For instance, it could be of order $m_\nu/M_1 \simlt 10^{-10}$. Thus, we expect that a decaying gravitino near the keV scale will have a sufficiently long lifetime as to be unobservable.

\section{Dark Matter Annihilation Scenarios}

If the signal comes from dark matter annihilation to two photons (with a dark matter mass of 3.5 keV), then we should fit it with a cross section $\left<\sigma v\right>_{\rm fit}$ that is related to the best-fit decay width $\Gamma_{\rm fit}$ studied above by
\beq
\int_{\rm l.o.s.} d\ell~n_{\rm DM} \Gamma_{\rm fit} = \int_{\rm l.o.s.} d\ell~n_{\rm DM}^2 \left<\sigma v\right>,
\eeq
where the integrals are along the line of sight. To obtain a first crude estimate, we read from Table 4 of Ref.~\cite{Bulbul:2014sua} that the Centaurus observations involved a total mass of $6.3\times 10^{13} M_\odot$ in a region of radius $R_{\rm ext} = 0.17~{\rm Mpc}$. If we treat the number density as constant over a spherical region (of course, it is not), we obtain an estimate:
\beq
n_{\rm DM} \sim \frac{6.3 \times 10^{13} M_\odot}{3.5~{\rm keV}} \frac{3}{4 \pi R_{\rm ext}^3} \approx 3 \times 10^4~{\rm cm}^{-3},
\eeq
leading to
\beq
\left<\sigma v\right>_{\rm fit} \approx 2.4 \times 10^{-33}~{\rm cm}^3/{\rm s} \approx 2 \times 10^{-16}~{\rm GeV}^{-2}.
\eeq
A more careful estimate could use an NFW profile, but given the number of sources involved, the approximate nature of the flux estimate, and the negative conclusion we will shortly reach about models, we will forgo this extra effort. The estimate for the flux from Perseus in terms of $\left<\sigma v\right>$ quoted in Ref.~\cite{Finkbeiner:2014sja} leads to an desired cross section of $10^{-32}~{\rm cm}^3/{\rm s}$, not far from our estimate above.

Suppose that dark matter is a fermion coupling to photons through the Rayleigh operator $\frac{1}{4\Lambda_R^3} {\bar \psi}\psi F^{\mu \nu}F_{\mu \nu}$. Then it has an annihilation rate~\cite{Weiner:2012cb}
\beq
\sigma v(\psi \psi \to \gamma\gamma) = \frac{m_\psi^4}{4 \pi \Lambda_R^6}.
\eeq
In this case, we fit the desired rate if:
\beq
\Lambda_R \approx 60~{\rm MeV}.
\eeq
Given that we know all charged particles (barring exotic millicharged particles, which would not couple strongly enough to generate this operator) with a mass below about 100 GeV, it seems impossible to UV complete this operator and achieve a large enough rate.

We could similarly consider the case of dark matter as a complex scalar coupling through the operator $\frac{1}{\Lambda_s^2} \phi^\dagger \phi F_{\mu \nu} F^{\mu \nu}$. This gives an annihilation cross section~\cite{Rajaraman:2012fu} 
\beq
\sigma v(\phi^\dagger \phi \to \gamma \gamma) = \frac{2 m_\phi^2}{\pi \Lambda_s^4},
\eeq
requiring
\beq
\Lambda_s \approx 14~{\rm GeV}.
\eeq
This is only marginally more plausible than the previous case. We conclude that, given the observed rate, it is unlikely that the X-ray line is accounted for by dark matter {\em directly} annihilating to a pair of photons.

\section{Discussion}

We have systematically examined the small set of operators involving dark matter decaying or annihilating directly to Standard Model particles and producing a photon line in the keV energy range. These are not all of the possibilities for generating such a line. For example, dark matter could annihilate or scatter to particles outside the Standard Model which subsequently decay. One such example has been presented in ref.~\cite{Finkbeiner:2014sja}, with dark matter self-interactions leading to collisional excitation and subsequent decay to the ground state through a dipole operator. This model has attractive features, including a self-interaction rate that could potentially account for small-scale structure puzzles like the lack of cusps in dwarf galaxy halos. 

So far the data on a possible 3.5 keV line have mostly been interpreted in the context of sterile neutrinos~\cite{Bulbul:2014sua,Boyarsky:2014jta,Ishida:2014dlp} or axion-like particles~\cite{Higaki:2014zua,Jaeckel:2014qea,Lee2014}. For this reason, we have considered the interpretation in terms of a scalar field---resembling the moduli fields predicted by string theory---in most detail. In particular, we have highlighted that without any theoretical bias, such an interpretation leads purely from bottom-up effective field theory considerations to a strong argument for additional new physics below 1000 TeV. Because such physics has to regulate a quartic divergence from a nonrenormalizable operator, the possible UV completions are limited. A composite theory could work in principle, but seems awkward to construct in practice. The most viable completion is supersymmetry, broken at or below the PeV-scale. This is an exciting possibility, with the relatively low energy scale in X-rays being linked to a scale probed by current or future colliders.

Even if the hint of a signal at 3.5 keV does not turn out to be related to dark matter, the analysis here would be valid for any future dark matter signals appearing in a similar energy range. Sterile neutrinos will remain a viable possibility, but we should also consider moduli and axion-like particles, which can offer tantalizing links between observational signals and very high energy scales.

\section*{Acknowledgments}

We thank Joe Conlon for useful correspondence on LARGE volume models, and Sam McDermott and Asher Berlin for spotting a factor of two error in the first version. MR thanks Marat Freytsis for a useful conversation. We thank JiJi Fan for comments on the draft. RK is supported by the Department of Energy Office of Science Graduate Fellowship Program (DOE SCGF), made possible in part by the American Recovery and Reinvestment Act of 2009, administered by ORISE-ORAU under contract no. DE-AC05-06OR23100.

\end{document}